\documentclass[aps,apl,twocolumn,groupedaddress,showpacs,amsmath]{revtex4}
\usepackage[latin1]{inputenc}    
\usepackage{graphicx}   

\begin{document}
\title{Measurements of higher order noise correlations in a quantum dot\\ with a finite bandwidth detector}
\author{S.~Gustavsson}
\email{simongus@phys.ethz.ch}
 \author{R.~Leturcq}
 \author{T.~Ihn}
 \author{K.~Ensslin}
 \affiliation {Solid State Physics Laboratory, ETH Z\"urich, CH-8093 Z\"urich,
 Switzerland}
 \author{M.~Reinwald}
 \author{W.~Wegscheider}
 \affiliation {Institut für experimentelle und angewandte Physik, Universität Regensburg, Germany}

\date{\today}

\begin{abstract}
We present measurements of the fourth and fifth cumulants of the
distribution of transmitted charge in a tunable quantum dot. We
investigate how the measured statistics is influenced by the finite
bandwidth of the detector and by the finite measurement time. By
including the detector when modeling the system, we use the theory
of full counting statistics to calculate the noise levels for the
combined system. The predictions of the finite-bandwidth model are
in good agreement with measured data.
\end{abstract}

\maketitle

Current fluctuations in mesoscopic systems have been extensively
studied due to the extra information they give in comparison to
measurements of the mean current \cite{blanter:00}. The focus has
traditionally been on investigations of the shot-noise, which for
classical systems arises due to the discreteness of the electron
charge. The theory of full counting statistics (FCS) was introduced
as a new way of examining current fluctuations \cite{levitov:96}.
With the FCS, fluctuations are studied by counting the number of
electrons that pass through a conductor within a fixed period of
time. This gives direct access to the probability distribution
function $p_{t_0}(N)$, which is the probability that $N$ electrons
are transferred within a time interval of length $t_0$. From the
distribution function, not only the shot noise but also correlations
of higher order can be calculated.

The third moment of a tunneling current has been shown to be
independent of the thermal noise \cite{levitovThird:2004,
fujisawa:2006}, thus making it a potential tool for investigating
electron-electron interactions even at elevated temperatures. Higher
order moments in strongly interacting systems are predicted to
depend heavily on both the conductance \cite{bagretsDiff:06} and on
the internal level structure \cite{belzig:05} of the system.
Determining higher order moments may therefore give a more complete
characterization of the electron transport process. This can be of
importance for realizing measurements of electron correlation and
entanglement effects in quantum dots \cite{loss:00, saraga:03}.


Experimentally, the third moment of the current distribution
function has been measured for a tunnel junction \cite{ReuletBomze}
as well as for a single quantum dot (QD)
\cite{gustavsson:05,gustavsson:2006} and a double QD
\cite{fujisawa:2006}.
In quantum optics, higher order moments are routinely measured in
order to study entanglement and coherence effects of the
electromagnetic field \cite{mandel:1995}. Here, we set out to
measure the fourth and fifth cumulant of the distribution function
for charge transport through a QD.

In general, experimental measurements of FCS for electrons are
difficult to achieve due to the need of a sensitive, high-bandwidth
detector capable of resolving individual electrons \cite{LuW:03,
fujisawa:04, bylander:05}. However, a more fundamental complication
with the measurements is that most forms of the FCS theory assume
the existence of (1) a detector with infinite bandwidth and (2)
infinitely long data traces. Since no physical detector can fulfill
these requirements, every experimental realization of the FCS will
measure a distribution which is influenced by the properties of the
detector. Here, we investigate how the violation of the two
assumptions modifies the measured statistics. By including the
detector in the model, the FCS for the combined QD-detector system
can be calculated. This model can explain the results for higher
order cumulants measured with a finite bandwidth detector.



\begin{figure}[htb]
\centering
 \includegraphics[width=\columnwidth]{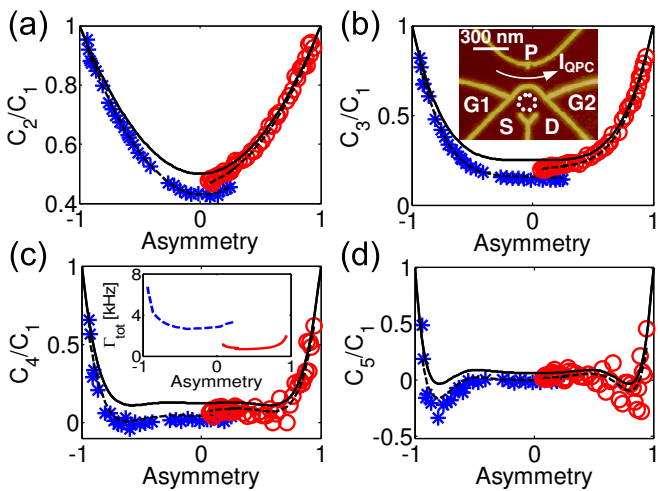}
 \caption{(a-d) Normalized cumulants $C_n/C_1$ versus dot asymmetry,
 $a = (\Gamma_{\mathrm{in}}-\Gamma_{\mathrm{out}})/(\Gamma_{\mathrm{in}}+\Gamma_{\mathrm{out}})$.
 The solid lines are theoretical predictions assuming a perfect
 detector, $C_2/C_1 = (1+a^2)/2$, $C_3/C_1 = (1+3a^4)/4$,
 $C_4/C_1 = (1+a^2-9 a^4+15 a^6)/8$ and $C_5/C_1 = (1+30 a^4-120 a^6+105 a^8)/16$.
 The dashed lines show the cumulants calculated from the
 model defined by Eq. (\ref{eq:mainM}) in the text.
 The inset in (b) shows the quantum dot with integrated charge read-out used in the experiment.
 The inset in (c) shows the variation of the total tunneling rate
 $\Gamma_{\mathrm{tot}} = \Gamma_{\mathrm{in}}
 +\Gamma_{\mathrm{out}}$ for the different measurement points.
 } \label{fig:CvsA}
\end{figure}

The sample consists of a QD [dotted circle in the inset of Fig.
\ref{fig:CvsA}(b)] with a nearby quantum point contact (QPC) used
for reading out the charge state of the QD \cite{field:93}. The
structure was fabricated using scanning probe litho\-graphy
\cite{fuhrer:04} on a $\mathrm{GaAs/Al_{0.3}Ga_{0.7}As}$
heterostructure with a two-dimensional electron gas 34 nm below the
surface. The gates $G_1$ and $G_2$ are used to tune the height of
the tunneling barriers connecting the dot to source and drain leads,
while the $P$-gate is used to keep the conductance through the QPC
in a regime where the sensitivity to changes in its electrostatic
potential is maximal.
All measurements were performed in a dilution refrigerator with a
base temperature of 20 mK. The electronic temperature extracted from
the width of Coulomb blockade resonances measured in the low bias
regime \cite{kouw:97} was $190~\mathrm{mK}$.

Due to the electrostatic coupling between the QD and the QPC, the
addition of an extra electron on the dot will cause a change in the
QPC conductance. By performing time-resolved measurements of the
current through the QPC, tunneling of single electrons can be
detected in real-time \cite{schl:04, vand:04, fujisawa:04}. In this
experiment, the QPC was voltage biased with $V_{QPC} = 250~\mu V$.
The current signal was sampled at $100~\mathrm{kHz}$, software
filtered at $4~\mathrm{kHz}$ using a 8th order Butterworth filter
and finally resampled at $20~\mathrm{kHz}$ to keep the amount of
data manageable.


To measure the higher cumulants for the current through the QD, one
has to generate the experimental probability density function
$p_{t_0}(N)$. This is done by splitting a time trace of length $T$
into $m=T/t_0$ intervals and counting the number of electrons
entering the dot within each interval. The cumulants are then
calculated directly from the distribution function. In a previous
experiment on a similar system the second and third cumulants were
measured \cite{gustavsson:05}. To extend the analysis to higher
cumulants, it is necessary to increase the length of the time traces
in order to collect more statistics. Here, we present cumulants
extracted from time traces of length $T=10~\mathrm{minutes}$. The
quality of the data allows to measure up to the fifth cumulant,
which is two orders higher than reported in previous experiments
\cite{ReuletBomze, gustavsson:05, fujisawa:2006}.


The results are shown in Fig. \ref{fig:CvsA}, where we plot the
normalized cumulants for different values of the asymmetry of the
tunneling rates,
$a=(\Gamma_{\mathrm{in}}-\Gamma_{\mathrm{out}})/(\Gamma_{\mathrm{in}}+\Gamma_{\mathrm{out}})$.
Here, $\Gamma_{\mathrm{in}}$ and $\Gamma_{\mathrm{out}}$ are the
rates for tunneling into and out of the dot, respectively. The
asymmetry is tuned by shifting the voltage on gate $G_1$ by an
amount $\Delta V$ and at the same time applying a compensating
voltage $-\Delta V$ on gate $G_2$. With the two gates having a
similar lever arm on the dot, the electrochemical potential of the
QD remains at the same level, but the height of the tunneling
barriers between the dot and the source and drain leads will change.
Doing so, we could tune the asymmetry from $a=-0.94$ to $a=+0.25$
while still keeping both tunneling rates within the measurement
bandwidth and avoiding charge rearrangements. To get data for the
full range of asymmetry, we did a second measurement at a different
gate voltage configuration. For the second set of data, the
asymmetry was tuned from $a=0.07$ to $a=0.93$. The stars and the
circles in Fig. \ref{fig:CvsA} represent data from the two different
sets of measurements. The measurements were performed with a QD bias
of $V_{\mathrm{bias}}=2.5~\mathrm{mV}$, with the electrochemical
potential of the dot far away from the Fermi levels of the source
and drain leads. This is to ensure that tunneling due to thermal
fluctuations is sufficiently suppressed \cite{gustavsson:05}.


The solid lines in Fig. \ref{fig:CvsA} depict the theoretical
predictions calculated from a two-state model \cite{bagrets:03}. The
analytical expressions are given in the figure caption. The higher
cumulants show a complex behavior as a function of the asymmetry,
with local minima at $a = \pm 0.6$ for $C_4/C_1$ and at $a = \pm
0.8$ for $C_5/C_1$. The fifth cumulant even becomes negative for
some configurations. The experimental data qualitatively agrees with
the theory, but for small values of the asymmetry there are
deviations from the expected behavior. The deviations are stronger
for the first set of data (stars). Since the tunneling rates in the
first measurement was about a factor of three higher than in the
second measurement [see inset of Fig. \ref{fig:CvsA}(c)], we suspect
the finite bandwidth of the detector to be a possible reason for the
discrepancies.

%
%
%


Recently, Naaman et al. \cite{naaman:2006} pointed out that
measurements of the transition rates of a Poisson two-state system
using a finite bandwidth detector always leads to an underestimate
of the rates. To determine the rates correctly, the detection rate
$\Gamma_{\mathrm{det}}$ of the detector must be taken into account.
In the low-bias, weak coupling Coulomb blockade regime, the QD can
be modeled as a Poisson two-state system. The two states correspond
to zero or one excess electron on the dot, and the transitions
between the two states occur whenever an electron tunnels into or
out of the dot. The probability distribution for the times needed
for an electron to tunnel into or out of the QD follows the
exponential $p_{\mathrm{in/out}}(t) = \Gamma_{\mathrm{in/out}}
\exp(-\Gamma_{\mathrm{in/out}}\,t )$ \cite{schl:04}.

\begin{figure}[htb]
\centering
 \includegraphics[width=\columnwidth]{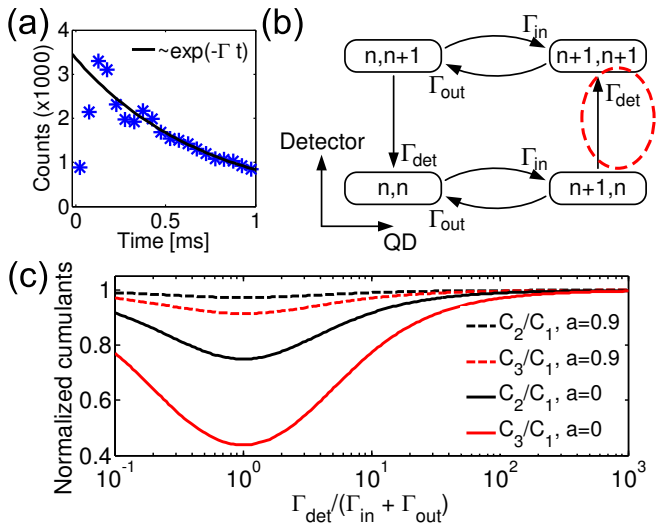}
 \caption{
 (a) Probability
 density of time needed for an electron to tunnel into the dot. Note
 the sharp decrease in counts for $t<100~\mathrm{\mu s}$ due to the
 finite bandwidth of the detector.
 The black curve is a long-time exponential fit with $\Gamma = 1.39~\mathrm{kHz}$.
 (b) Model for the dot-detector
 system. A state $(n,m)$ corresponds to $n$ electrons on the dot
 while the detector at the same time is measuring $m$ electrons.
 (c) Higher cumulants versus relative detection bandwidth
 $\Gamma_{\mathrm{det}}/(\Gamma_{\mathrm{in}}+\Gamma_{\mathrm{out}})$,
 calculated from the model in
 (b). The cumulants are normalized to the results from the infinite-bandwidth case. The
 influence of the finite bandwidth is maximal when the asymmetry
 $a=(\Gamma_{\mathrm{in}}-\Gamma_{\mathrm{out}})/(\Gamma_{\mathrm{in}}+\Gamma_{\mathrm{out}})$
 is zero.
 } \label{fig:model}
\end{figure}

An example of a probability distribution taken from measured data is
shown in Fig. \ref{fig:model}(a). The long-time behavior is
exponential, but for times $t<100~\mathrm{\mu s}$ there is a sharp
decrease in the number of counts registered by the detector. From
the figure, we can estimate $\tau_{\mathrm{det}}$, which is the
average time it takes for the detector to register an event. We find
$\tau_{\mathrm{det}} = 70~\mathrm{\mu s}$, giving a detection rate
of $\Gamma_{\mathrm{det}} = 1/\tau_{\mathrm{det}} =
14~\mathrm{kHz}$. Note that the detection rate
$\Gamma_{\mathrm{det}}$ does not only depend on the measurement
bandwidth but also on the signal-to-noise ratio of the detector
signal as well as the redundancy needed to minimize the risk of
detecting false events \cite{shannon:1949}. All tunneling rates
presented in the following have been extracted from distributions
such as the one shown in Fig. \ref{fig:model}(a), using the methods
described in Ref. \cite{naaman:2006} with
$\Gamma_{\mathrm{det}}=14~\mathrm{kHz}$.


The finite bandwidth will also influence the FCS measured by the
detector. Following the ideas of Ref. \cite{naaman:2006}, we account
for the finite bandwidth by including the states of the detector in
the model. Figure \ref{fig:model}(b) shows the four possible states
of the combined dot-detector model. The state $(n+1,n)$ refers to a
situation where there are $n+1$ electrons on the dot, while the
detector at the same time reads $n$ electrons. The transition from
the state $(n+1,n)$ to the state $(n+1,n+1)$ occurs when the
detector registers the electron. This process occurs with the rate
of the detector, $\Gamma_{\mathrm{det}}$.

To calculate the FCS for the QD-detector system, we write the master
equation $\dot{P} = M\,P$, with $P =
[(n,n),(n+1,n),(n,n+1),(n+1,n+1)]$ and
\begin{equation}\label{eq:mainM}
M_{\chi}=\left(
\begin{array}{cccc}
  -\Gamma_{\mathrm{in}} & \Gamma_{\mathrm{out}} & \Gamma_{\mathrm{det}} & 0 \\
  \Gamma_{\mathrm{in}} & -(\Gamma_{\mathrm{out}}+\Gamma_{\mathrm{det}}) & 0 & 0 \\
  0 & 0 & -(\Gamma_{\mathrm{in}} + \Gamma_{\mathrm{det}}) & \Gamma_{\mathrm{out}} \\
  0 & \Gamma_{\mathrm{det}}* e^{i\chi} & \Gamma_{\mathrm{in}} &
  -\Gamma_{\mathrm{out}}
\end{array}
\right).
\end{equation}
In the above matrix, we have included the counting factor
$e^{i\chi}$ at the element where the detector registers an electron
tunneling into the dot [see dashed circle in Fig.
\ref{fig:model}(b)]. The statistics obtained in this way relates
directly to what is measured in the experiment. Using the methods of
Ref. \cite{bagrets:03}, we calculate the first few cumulants for the
above expression as a function of relative bandwidth $k =
\Gamma_\mathrm{det}/(\Gamma_{\mathrm{in}} + \Gamma_{\mathrm{out}})$
and asymmetry $a = (\Gamma_{\mathrm{in}} -
\Gamma_{\mathrm{out}})/(\Gamma_{\mathrm{in}} +
\Gamma_{\mathrm{out}})$. The normalized second and third cumulants
take the form
\begin{eqnarray}
 \label{eq:C2BW}
 C_2/C_1 &=&
 \frac{1+a^2}{2} - \frac{k (1-a^2)}{2(1+k)^2},\\
 \label{eq:C3BW}
 C_3/C_1 &=&
 \frac{1+ 3a^4}{4} -
 \frac{3 k (1 + k + k^2 )}{4(1+k)^4} - \nonumber\\
 & & \frac{6 \, a^2 k^2}{4(1+k)^4} + \frac{3\, a^4 k (1 + 3 k + k^2
 )}{4(1+k)^4}.
\end{eqnarray}



In Fig. \ref{fig:model}(c) we plot the second and third cumulants
from Eq. (\ref{eq:C2BW}) and Eq. (\ref{eq:C3BW}) for different
values of asymmetry $a$ and relative bandwidth $k$. The cumulants
have been normalized to the values for the infinite bandwidth
detector. With $\Gamma_\mathrm{det} \gg \Gamma_{\mathrm{in}} +
\Gamma_{\mathrm{out}}$, the cumulants approach the infinite
bandwidth result, as expected. However, even with
$\Gamma_\mathrm{det} = 10(\Gamma_{\mathrm{in}} +
\Gamma_{\mathrm{out}})$ and perfect symmetry ($a=0$), the second
cumulant deviates by almost 10\% and the third cumulant by more than
20\% from the perfect detector values. As the bandwidth is further
decreased, the deviations grow stronger and reach a maximum as
$\Gamma_\mathrm{det} = \Gamma_{\mathrm{in}} +
\Gamma_{\mathrm{out}}$. With $\Gamma_\mathrm{det} \ll
\Gamma_{\mathrm{in}} + \Gamma_{\mathrm{out}}$, the cumulants once
again approach the perfect detector values. When the detector is
much slower than the underlying tunneling process, it will only
sample the average population of the two states. In this limit, the
dynamics of the system does not interfere with the dynamics of the
detector and we recover the correct relative noise levels. It should
be noted that this is true only for the noise relative to the
detected mean current. Since the detector will miss most of the
tunneling events, the absolute values of both the current and the
noise will be underestimated.

We have also performed the analysis for the fourth and fifth
cumulants. We do not show the analytical expressions due to space
limitations; however, the results corresponding to the experimental
configuration are represented by the dashed lines in Fig. 1. Over
the full range of bandwidth and asymmetry, we find that the noise
detected with the finite bandwidth system is always lower than for
the ideal detector case. The reduction can be qualitatively
understood by considering the probability distribution $p_{t_0}(N)$.
The finite bandwidth makes it less probable to detect fast events,
meaning that the probability of detecting a large number of
electrons within the interval $t_0$ will decrease more than the
probability of detecting few electrons. This will cut the high-count
tail of the distribution and thereby reduce its width ($C_2$) and
its skewness ($C_3$). An interesting feature is that the cumulants
calculated for a less symmetric configurations [$a=0.9$ in Fig.
\ref{fig:model}(c)] show less influence of the
finite bandwidth. 

A second limitation of a general FCS measurement is the finite
length of each time trace. In order to generate the experimental
probability density function $p_{t_0}(N)$, the total trace of length
$T$ must be split into $m=T/t_0$ intervals, each of length $t_0$.
Most FCS theories only predict results for the case $t_0 \gg
1/\Gamma$, where $\Gamma$ is a typical transition rate of the
system.
In the experiment, it is favorable to make $t_0$ as short as
possible in order to increase the number of samples $m=T/t_0$. This
will improve the quality of the distribution and help to minimize
statistical errors.

The condition $t_0 \gg \Gamma$ is imposed by the approximation that
the cumulant generating function (CGF) $S(\chi)$ for $p_{t_0}(N)$
only depends on the lowest eigenvalue $\Lambda_{\mathrm{min}}$ of
the master equation matrix $M_{\chi}$, with $S(\chi) = -t_0
\Lambda_{\mathrm{min}}$. A FCS valid for finite $t_0$ must include
all eigenvalues and eigenvectors of $M_{\chi}$ \cite{bagrets:03} .
The corresponding expression is
\begin{equation}\label{eq:SAll}
    \exp[S(\chi)] = \langle q_0 | p^{(n)}\rangle
    \exp(-t_0 \Lambda_n)
    \langle q^{(n)} | p_0\rangle,
\end{equation}
where $\langle q^{(n)}|$ and $|p^{(n)}\rangle$ are the left and
right eigenvectors of the matrix $M_{\chi}$, $\Lambda_{n}$ are the
eigenvalues of $M_{\chi}$ and $\langle q_0|,~|p_0\rangle$ are the
eigenvectors corresponding to the lowest eigenvalue
$\Lambda_{\mathrm{min}}$. The cumulants generated from the CGF in
Eq. (\ref{eq:SAll}) will in general be a function of $t_0$.

\begin{figure}[htb]
\centering
 \includegraphics[width=\columnwidth]{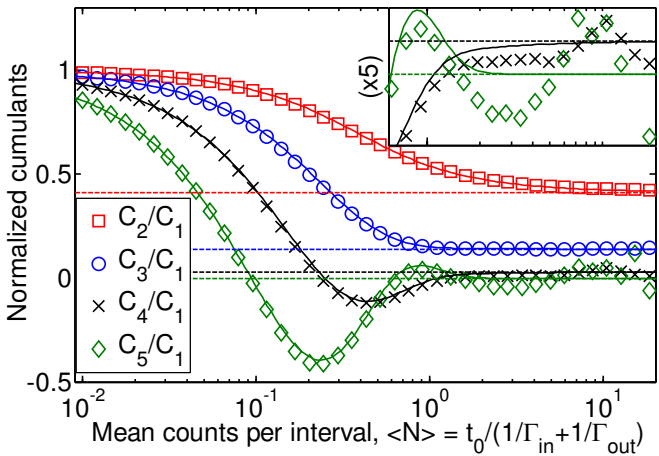}
 \caption{Normalized cumulants evaluated for different lengths of the time
 interval $t_0$. The symbols show the experimental
 data, extracted from a time trace of length $T=10~\mathrm{minutes}$,
 containing 350595 events, with $a= 0.053$, and $\Gamma_{\mathrm{tot}} =
 3062~\mathrm{Hz}$. The solid lines are calculations from the
 FCS given by Eq. (\ref{eq:SAll}) in the text, while the
 dashed lines are the asymptotes for $t_0 \rightarrow \infty$. The
 inset shows a magnification of the vertical axis (horizontal
 axis unchanged) for $C_4/C_1$ and $C_5/C_1$ for $\langle N \rangle >
 0.6$.
 }
\label{fig:CvsN}
\end{figure}

To investigate how small $t_0$ can be before systematic errors
become relevant, we calculate the cumulants from the CGF of Eq.
(\ref{eq:SAll}) with the master equation matrix $M_{\chi}$ of Eq.
(\ref{eq:mainM}). The results are shown in Fig. \ref{fig:CvsN},
where we plot the normalized cumulants as a function of the mean
number of counts per interval, $\langle N \rangle =
t_0/(1/\Gamma_{\mathrm{in}} + 1/\Gamma_{\mathrm{out}})$. The symbols
show cumulants extracted from measured data ($T =
10~\mathrm{minutes}$, $a=0.053$, $\Gamma_{\mathrm{in}} +
\Gamma_{\mathrm{out}} = 3062~\mathrm{Hz}$ and
$\Gamma_{\mathrm{det}}=14~\mathrm{kHz}$), while the solid lines are
results from the CGF for the same set of parameters. The dashed
lines are the asymptotes for the limiting case $t_0 \rightarrow
\infty$.

In general, the data and the theory are in good agreement. There are
some deviations in the fourth and fifth cumulants for large $t_0$
($\langle N \rangle>6$ in Fig. \ref{fig:CvsN}), but these are
statistical errors due to the shortness of the total time trace. For
short $t_0$, all cumulants converge to $C_n/C_1 \rightarrow 1$. This
is because as $\langle N \rangle \ll 1$, the probability
distribution $p_{t_0}(N)$ will be non-zero only for $N=0$ and $N=1$,
with $p_{t_0}(0)=1-q$, $p_{t_0}(1)=q$ and $q=\langle N \rangle$.
This is the definition of a Bernoulli distribution, for which the
normalized cumulants $C_n/C_1 \rightarrow 1$ as $q\rightarrow 0$
\cite{mathWorldBernoulli}.

Focusing on the other regime, $\langle N \rangle > 1$, we see that
cumulants of different orders converge to their asymptotic limits
for different values of $t_0$. The second cumulant needs a longer
interval $t_0$ to reach a specified tolerance compared to the higher
cumulants. This is of interest for the experimental determination of
higher cumulants.
By choosing a shorter value of $t_0$ when calculating higher
cumulants, the amount of samples $m = T/t_0$ can be increased. For
the data in Fig. \ref{fig:CvsA}, the cumulants were calculated with
intervals $t_0$ giving $\langle N \rangle = 15$ for $C_2$, $\langle
N \rangle = 6$ for $C_3$, $\langle N \rangle = 3$ for $C_4$ and
$\langle N \rangle = 2$ for $C_5$.
The maximal deviations between the correct cumulants and the ones
determined with a finite length $t_0$ can be estimated by checking
the convergence for all values of the asymmetry. For the data shown
in Fig. \ref{fig:CvsA}, we find $\Delta C_2/C_1 = 0.007$, $\Delta
C_3/C_1 = 0.009$, $\Delta C_4/C_1 = 0.01$ and $\Delta C_5/C_1 =
0.03$.


Coming back to the results of Fig. \ref{fig:CvsA}, we are now able
explain why the measured cumulants show lower values compared to the
perfect-detector theory.
The dashed lines in Fig. \ref{fig:CvsA} are the cumulants calculated
from the combined QD-detector model of Eq. (\ref{eq:mainM}), with
$\Gamma_{\mathrm{det}} = 14~\mathrm{kHz}$. The overall agreement is
good, especially since no fitting parameters are involved. Higher
cumulants end up to be slightly lower than theory predicts. We
speculate that the deviations could be due to low-frequency
fluctuations of the tunneling rates over the time of measurement.


In conclusion, we have measured the first five cumulants of the
distribution of charge transmitted through a QD.
The ability to measure higher cumulants shows that we can determine
the distribution function very precisely. The high accuracy of the
technique makes it a promising tool for probing subtle effects in
the transport statistics of more complex QD systems.
We have found that the measured statistics depends strongly on the
bandwidth of the charge detector. By including the detector in the
model, we show that the framework of FCS can be used to predict
noise levels for systems with a finite bandwidth detector. The
principle is general and can be applied to any rate-equation model
used for calculating the FCS. Financial support by the NCCR
Nanoscience through the Swiss Science Foundation (Schweizerischer
Nationalfonds) is gratefully acknowledged.


\bibliographystyle{apsrev}
\bibliography{FiniteBW}

\begin{thebibliography}{23}
\expandafter\ifx\csname natexlab\endcsname\relax\def\natexlab#1{#1}\fi
\expandafter\ifx\csname bibnamefont\endcsname\relax
  \def\bibnamefont#1{#1}\fi
\expandafter\ifx\csname bibfnamefont\endcsname\relax
  \def\bibfnamefont#1{#1}\fi
\expandafter\ifx\csname citenamefont\endcsname\relax
  \def\citenamefont#1{#1}\fi
\expandafter\ifx\csname url\endcsname\relax
  \def\url#1{\texttt{#1}}\fi
\expandafter\ifx\csname urlprefix\endcsname\relax\def\urlprefix{URL }\fi
\providecommand{\bibinfo}[2]{#2}
\providecommand{\eprint}[2][]{\url{#2}}

\bibitem[{\citenamefont{Blanter and B\"uttiker}(2000)}]{blanter:00}
\bibinfo{author}{\bibfnamefont{Y.~M.} \bibnamefont{Blanter}} \bibnamefont{and}
  \bibinfo{author}{\bibfnamefont{M.}~\bibnamefont{B\"uttiker}},
  \bibinfo{journal}{Physics Reports} \textbf{\bibinfo{volume}{336}},
  \bibinfo{pages}{1} (\bibinfo{year}{2000}).

\bibitem[{\citenamefont{Levitov et~al.}(1996)\citenamefont{Levitov, Lee, and
  Lesovik}}]{levitov:96}
\bibinfo{author}{\bibfnamefont{L.~S.} \bibnamefont{Levitov}},
  \bibinfo{author}{\bibfnamefont{H.~W.} \bibnamefont{Lee}}, \bibnamefont{and}
  \bibinfo{author}{\bibfnamefont{G.~B.} \bibnamefont{Lesovik}},
  \bibinfo{journal}{J. Math. Phys.} \textbf{\bibinfo{volume}{37}},
  \bibinfo{pages}{4845} (\bibinfo{year}{1996}).

\bibitem[{\citenamefont{Fujisawa et~al.}(2006)\citenamefont{Fujisawa, Hayashi,
  Tomita, and Hirayama}}]{fujisawa:2006}
\bibinfo{author}{\bibfnamefont{T.}~\bibnamefont{Fujisawa}},
  \bibinfo{author}{\bibfnamefont{T.}~\bibnamefont{Hayashi}},
  \bibinfo{author}{\bibfnamefont{R.}~\bibnamefont{Tomita}}, \bibnamefont{and}
  \bibinfo{author}{\bibfnamefont{Y.}~\bibnamefont{Hirayama}},
  \bibinfo{journal}{Science} \textbf{\bibinfo{volume}{312}},
  \bibinfo{pages}{1634} (\bibinfo{year}{2006}).

\bibitem[{\citenamefont{Levitov and Reznikov}(2004)}]{levitovThird:2004}
\bibinfo{author}{\bibfnamefont{L.~S.} \bibnamefont{Levitov}} \bibnamefont{and}
  \bibinfo{author}{\bibfnamefont{M.}~\bibnamefont{Reznikov}},
  \bibinfo{journal}{Phys. Rev. B} \textbf{\bibinfo{volume}{70}},
  \bibinfo{pages}{115305} (\bibinfo{year}{2004}).

\bibitem[{\citenamefont{Bagrets et~al.}(2006)\citenamefont{Bagrets, Utsumi,
  Golubev, and Schön}}]{bagretsDiff:06}
\bibinfo{author}{\bibfnamefont{D.}~\bibnamefont{Bagrets}},
  \bibinfo{author}{\bibfnamefont{Y.}~\bibnamefont{Utsumi}},
  \bibinfo{author}{\bibfnamefont{D.}~\bibnamefont{Golubev}}, \bibnamefont{and}
  \bibinfo{author}{\bibfnamefont{G.}~\bibnamefont{Schön}},
  \bibinfo{journal}{Fortschritte der Physik 54, 917-938 (2006).}
  \textbf{\bibinfo{volume}{54}}, \bibinfo{pages}{917} (\bibinfo{year}{2006}).

\bibitem[{\citenamefont{Belzig}(2005)}]{belzig:05}
\bibinfo{author}{\bibfnamefont{W.}~\bibnamefont{Belzig}},
  \bibinfo{journal}{Phys. Rev. B} \textbf{\bibinfo{volume}{71}},
  \bibinfo{pages}{161301(R)} (\bibinfo{year}{2005}).

\bibitem[{\citenamefont{Loss and Sukhorukov}(2000)}]{loss:00}
\bibinfo{author}{\bibfnamefont{D.}~\bibnamefont{Loss}} \bibnamefont{and}
  \bibinfo{author}{\bibfnamefont{E.~V.} \bibnamefont{Sukhorukov}},
  \bibinfo{journal}{Phys. Rev. Lett.} \textbf{\bibinfo{volume}{84}},
  \bibinfo{pages}{1035} (\bibinfo{year}{2000}).

\bibitem[{\citenamefont{Saraga and Loss}(2003)}]{saraga:03}
\bibinfo{author}{\bibfnamefont{D.~S.} \bibnamefont{Saraga}} \bibnamefont{and}
  \bibinfo{author}{\bibfnamefont{D.}~\bibnamefont{Loss}},
  \bibinfo{journal}{Phys. Rev. Lett.} \textbf{\bibinfo{volume}{90}},
  \bibinfo{pages}{166803} (\bibinfo{year}{2003}).

\bibitem[{\citenamefont{Reulet et~al.}(2003)\citenamefont{Reulet, Senzier, and
  Prober}}]{ReuletBomze}
\bibinfo{author}{\bibfnamefont{B.}~\bibnamefont{Reulet}},
  \bibinfo{author}{\bibfnamefont{J.}~\bibnamefont{Senzier}}, \bibnamefont{and}
  \bibinfo{author}{\bibfnamefont{D.~E.} \bibnamefont{Prober}},
  \bibinfo{journal}{Phys. Rev. Lett.} \textbf{\bibinfo{volume}{91}},
  \bibinfo{pages}{196601} (\bibinfo{year}{2003});
\bibinfo{author}{\bibfnamefont{Y.}~\bibnamefont{Bomze}} {\it et al.},
  \bibinfo{journal}{Phys. Rev. Lett.} \textbf{\bibinfo{volume}{95}},
  \bibinfo{pages}{176601} (\bibinfo{year}{2005}).

\bibitem[{\citenamefont{Gustavsson
  et~al.}(2006{\natexlab{a}})\citenamefont{Gustavsson, Leturcq, Simovic,
  Schleser, Ihn, Studerus, Ensslin, Driscoll, and Gossard}}]{gustavsson:05}
\bibinfo{author}{\bibfnamefont{S.}~\bibnamefont{Gustavsson}},
  \bibinfo{author}{\bibfnamefont{R.}~\bibnamefont{Leturcq}},
  \bibinfo{author}{\bibfnamefont{B.}~\bibnamefont{Simovic}},
  \bibinfo{author}{\bibfnamefont{R.}~\bibnamefont{Schleser}},
  \bibinfo{author}{\bibfnamefont{T.}~\bibnamefont{Ihn}},
  \bibinfo{author}{\bibfnamefont{P.}~\bibnamefont{Studerus}},
  \bibinfo{author}{\bibfnamefont{K.}~\bibnamefont{Ensslin}},
  \bibinfo{author}{\bibfnamefont{D.~C.} \bibnamefont{Driscoll}},
  \bibnamefont{and} \bibinfo{author}{\bibfnamefont{A.~C.}
  \bibnamefont{Gossard}}, \bibinfo{journal}{Phys. Rev. Lett.}
  \textbf{\bibinfo{volume}{96}}, \bibinfo{pages}{076605}
  (\bibinfo{year}{2006}{\natexlab{a}}).

\bibitem[{\citenamefont{Gustavsson
  et~al.}(2006{\natexlab{b}})\citenamefont{Gustavsson, Leturcq, Simovic,
  Schleser, Studerus, Ihn, Ensslin, Driscoll, and Gossard}}]{gustavsson:2006}
\bibinfo{author}{\bibfnamefont{S.}~\bibnamefont{Gustavsson}},
  \bibinfo{author}{\bibfnamefont{R.}~\bibnamefont{Leturcq}},
  \bibinfo{author}{\bibfnamefont{B.}~\bibnamefont{Simovic}},
  \bibinfo{author}{\bibfnamefont{R.}~\bibnamefont{Schleser}},
  \bibinfo{author}{\bibfnamefont{P.}~\bibnamefont{Studerus}},
  \bibinfo{author}{\bibfnamefont{T.}~\bibnamefont{Ihn}},
  \bibinfo{author}{\bibfnamefont{K.}~\bibnamefont{Ensslin}},
  \bibinfo{author}{\bibfnamefont{D.~C.} \bibnamefont{Driscoll}},
  \bibnamefont{and} \bibinfo{author}{\bibfnamefont{A.~C.}
  \bibnamefont{Gossard}}, \bibinfo{journal}{Phys. Rev. B}
  \textbf{\bibinfo{volume}{74}}, \bibinfo{pages}{195305}
  (\bibinfo{year}{2006}{\natexlab{b}}).

\bibitem[{\citenamefont{Mandel and Wolf}(1995)}]{mandel:1995}
\bibinfo{author}{\bibfnamefont{L.}~\bibnamefont{Mandel}} \bibnamefont{and}
  \bibinfo{author}{\bibfnamefont{E.}~\bibnamefont{Wolf}},
  \emph{\bibinfo{title}{Optical Coherence and Quantum Optics}}
  (\bibinfo{publisher}{Cambridge University Press}, \bibinfo{year}{1995}).

\bibitem[{\citenamefont{Lu et~al.}(2003)\citenamefont{Lu, Ji, Pfeiffer, West,
  and Rimberg}}]{LuW:03}
\bibinfo{author}{\bibfnamefont{W.}~\bibnamefont{Lu}},
  \bibinfo{author}{\bibfnamefont{Z.}~\bibnamefont{Ji}},
  \bibinfo{author}{\bibfnamefont{L.}~\bibnamefont{Pfeiffer}},
  \bibinfo{author}{\bibfnamefont{K.~W.} \bibnamefont{West}}, \bibnamefont{and}
  \bibinfo{author}{\bibfnamefont{A.~J.} \bibnamefont{Rimberg}},
  \bibinfo{journal}{Nature} \textbf{\bibinfo{volume}{423}},
  \bibinfo{pages}{422} (\bibinfo{year}{2003}).

\bibitem[{\citenamefont{Fujisawa et~al.}(2004)\citenamefont{Fujisawa, Hayashi,
  Hirayama, Cheong, and Jeong}}]{fujisawa:04}
\bibinfo{author}{\bibfnamefont{T.}~\bibnamefont{Fujisawa}},
  \bibinfo{author}{\bibfnamefont{T.}~\bibnamefont{Hayashi}},
  \bibinfo{author}{\bibfnamefont{Y.}~\bibnamefont{Hirayama}},
  \bibinfo{author}{\bibfnamefont{H.~D.} \bibnamefont{Cheong}},
  \bibnamefont{and} \bibinfo{author}{\bibfnamefont{Y.~H.} \bibnamefont{Jeong}},
  \bibinfo{journal}{Appl. Phys. Lett.} \textbf{\bibinfo{volume}{84}},
  \bibinfo{pages}{2343} (\bibinfo{year}{2004}).

\bibitem[{\citenamefont{Bylander et~al.}(2005)\citenamefont{Bylander, Duty, and
  Delsing}}]{bylander:05}
\bibinfo{author}{\bibfnamefont{J.}~\bibnamefont{Bylander}},
  \bibinfo{author}{\bibfnamefont{T.}~\bibnamefont{Duty}}, \bibnamefont{and}
  \bibinfo{author}{\bibfnamefont{P.}~\bibnamefont{Delsing}},
  \bibinfo{journal}{Nature} \textbf{\bibinfo{volume}{434}},
  \bibinfo{pages}{361} (\bibinfo{year}{2005}).

\bibitem[{\citenamefont{Field et~al.}(1993)\citenamefont{Field, Smith, Pepper,
  Ritchie, Frost, Jones, and Hasko}}]{field:93}
\bibinfo{author}{\bibfnamefont{M.}~\bibnamefont{Field}},
  \bibinfo{author}{\bibfnamefont{C.~G.} \bibnamefont{Smith}},
  \bibinfo{author}{\bibfnamefont{M.}~\bibnamefont{Pepper}},
  \bibinfo{author}{\bibfnamefont{D.~A.} \bibnamefont{Ritchie}},
  \bibinfo{author}{\bibfnamefont{J.~E.~F.} \bibnamefont{Frost}},
  \bibinfo{author}{\bibfnamefont{G.~A.~C.} \bibnamefont{Jones}},
  \bibnamefont{and} \bibinfo{author}{\bibfnamefont{D.~G.} \bibnamefont{Hasko}},
  \bibinfo{journal}{Phys. Rev. Lett.} \textbf{\bibinfo{volume}{70}},
  \bibinfo{pages}{1311} (\bibinfo{year}{1993}).

\bibitem[{\citenamefont{Fuhrer et~al.}(2002)\citenamefont{Fuhrer, Dorn,
  L\"uscher, Heinzel, Ensslin, Wegscheider, and Bichler}}]{fuhrer:04}
\bibinfo{author}{\bibfnamefont{A.}~\bibnamefont{Fuhrer}},
  \bibinfo{author}{\bibfnamefont{A.}~\bibnamefont{Dorn}},
  \bibinfo{author}{\bibfnamefont{S.}~\bibnamefont{L\"uscher}},
  \bibinfo{author}{\bibfnamefont{T.}~\bibnamefont{Heinzel}},
  \bibinfo{author}{\bibfnamefont{K.}~\bibnamefont{Ensslin}},
  \bibinfo{author}{\bibfnamefont{W.}~\bibnamefont{Wegscheider}},
  \bibnamefont{and} \bibinfo{author}{\bibfnamefont{M.}~\bibnamefont{Bichler}},
  \bibinfo{journal}{Superl. and Microstruc.} \textbf{\bibinfo{volume}{31}},
  \bibinfo{pages}{19} (\bibinfo{year}{2002}).

\bibitem[{\citenamefont{Kouwenhoven et~al.}(1997)\citenamefont{Kouwenhoven,
  Marcus, McEuen, Tarucha, Westervelt, and Wingreen}}]{kouw:97}
\bibinfo{author}{\bibfnamefont{L.~P.} \bibnamefont{Kouwenhoven}},
  \bibinfo{author}{\bibfnamefont{C.~M.} \bibnamefont{Marcus}},
  \bibinfo{author}{\bibfnamefont{P.~M.} \bibnamefont{McEuen}},
  \bibinfo{author}{\bibfnamefont{S.}~\bibnamefont{Tarucha}},
  \bibinfo{author}{\bibfnamefont{R.~M.} \bibnamefont{Westervelt}},
  \bibnamefont{and} \bibinfo{author}{\bibfnamefont{N.~S.}
  \bibnamefont{Wingreen}}, in \emph{\bibinfo{booktitle}{Mesoscopic Electron
  Transport}}, edited by \bibinfo{editor}{\bibfnamefont{L.~L.}
  \bibnamefont{Sohn}}, \bibinfo{editor}{\bibfnamefont{L.~P.}
  \bibnamefont{Kouwenhoven}}, \bibnamefont{and}
  \bibinfo{editor}{\bibfnamefont{G.}~\bibnamefont{Sch\"on}}
  (\bibinfo{publisher}{Kluwer}, \bibinfo{address}{Dordrecht},
  \bibinfo{year}{1997}), NATO ASI Ser. E 345, pp. \bibinfo{pages}{105--214}.

\bibitem[{\citenamefont{Schleser et~al.}(2004)\citenamefont{Schleser, Ruh, Ihn,
  Ensslin, Driscoll, and Gossard}}]{schl:04}
\bibinfo{author}{\bibfnamefont{R.}~\bibnamefont{Schleser}},
  \bibinfo{author}{\bibfnamefont{E.}~\bibnamefont{Ruh}},
  \bibinfo{author}{\bibfnamefont{T.}~\bibnamefont{Ihn}},
  \bibinfo{author}{\bibfnamefont{K.}~\bibnamefont{Ensslin}},
  \bibinfo{author}{\bibfnamefont{D.~C.} \bibnamefont{Driscoll}},
  \bibnamefont{and} \bibinfo{author}{\bibfnamefont{A.~C.}
  \bibnamefont{Gossard}}, \bibinfo{journal}{Appl. Phys. Lett.}
  \textbf{\bibinfo{volume}{85}}, \bibinfo{pages}{2005} (\bibinfo{year}{2004}).

\bibitem[{\citenamefont{Vandersypen et~al.}(2004)\citenamefont{Vandersypen,
  Elzerman, Schouten, Willems~van Beveren, Hanson, and Kouwenhoven}}]{vand:04}
\bibinfo{author}{\bibfnamefont{L.~M.~K.} \bibnamefont{Vandersypen}},
  \bibinfo{author}{\bibfnamefont{J.~M.} \bibnamefont{Elzerman}},
  \bibinfo{author}{\bibfnamefont{R.~N.} \bibnamefont{Schouten}},
  \bibinfo{author}{\bibfnamefont{L.~H.} \bibnamefont{Willems~van Beveren}},
  \bibinfo{author}{\bibfnamefont{R.}~\bibnamefont{Hanson}}, \bibnamefont{and}
  \bibinfo{author}{\bibfnamefont{L.~P.} \bibnamefont{Kouwenhoven}},
  \bibinfo{journal}{Appl. Phys. Lett.} \textbf{\bibinfo{volume}{85}},
  \bibinfo{pages}{4394} (\bibinfo{year}{2004}).

\bibitem[{\citenamefont{Bagrets and Nazarov}(2003)}]{bagrets:03}
\bibinfo{author}{\bibfnamefont{D.~A.} \bibnamefont{Bagrets}} \bibnamefont{and}
  \bibinfo{author}{\bibfnamefont{Y.~V.} \bibnamefont{Nazarov}},
  \bibinfo{journal}{Phys. Rev. B} \textbf{\bibinfo{volume}{67}},
  \bibinfo{pages}{085316} (\bibinfo{year}{2003}).

\bibitem[{\citenamefont{Naaman and Aumentado}(2006)}]{naaman:2006}
\bibinfo{author}{\bibfnamefont{O.}~\bibnamefont{Naaman}} \bibnamefont{and}
  \bibinfo{author}{\bibfnamefont{J.}~\bibnamefont{Aumentado}},
  \bibinfo{journal}{Phys. Rev. Lett.} \textbf{\bibinfo{volume}{96}},
  \bibinfo{pages}{100201} (\bibinfo{year}{2006}).

\bibitem[{\citenamefont{Shannon}(1949)}]{shannon:1949}
\bibinfo{author}{\bibfnamefont{C.~E.} \bibnamefont{Shannon}},
  \bibinfo{journal}{Proc. Institute of Radio Engineers}
  \textbf{\bibinfo{volume}{37}}, \bibinfo{pages}{10} (\bibinfo{year}{1949}).

\bibitem[{\citenamefont{Weisstein}()}]{mathWorldBernoulli}
\bibinfo{author}{\bibfnamefont{E.~W.} \bibnamefont{Weisstein}},
  \bibinfo{note}{"Bernoulli Distribution." From MathWorld - A Wolfram Web
  Resource. http://mathworld.wolfram.com/BernoulliDistribution.html}.

\end{thebibliography}

\end{document}